\documentclass[aps,preprint]{revtex4-1}
\usepackage{eurosym}
\usepackage{bm}
\usepackage{amsfonts}
\usepackage{amsmath}
\usepackage{amssymb}
\usepackage{graphicx}

\begin{document}

\title{Neutrino emission from Cooper pairs at finite temperatures}
\author{Lev B. Leinson}
\email{Correspondence should be addressed to Lev Leinson; leinson@yandex.ru}
\affiliation{Pushkov Institute of Terrestrial Magnetism, Ionosphere and 
Radiowave Propagation of the Russian Academy of Science (IZMIRAN),\\
108840 Troitsk, Moscow, Russia}

\begin{abstract}
A brief review is given of the current state of the problem of neutrino pair
emission through neutral weak currents caused by the Cooper pairs breaking
and formation (PBF) in superfluid baryon matter at thermal equilibrium. The
cases of singlet-state pairing with isotropic superfluid gap and
spin-triplet pairing with an anisotropic gap are analyzed with allowance for
the anomalous weak interactions caused by superfluidity. It is shown that
taking into account the anomalous weak interactions in both the vector and
axial channels is very important for a correct description of neutrino
energy losses through the PBF processes. The anomalous contributions lead to
an almost complete suppression of the PBF neutrino emission in spin-singlet
superfluids and strong reduction of the PBF neutrino losses in the
spin-triplet superfluid neutron matter, which considerably slows down the
cooling rate of neutron stars with superfluid cores.
\end{abstract}

\maketitle

%%%%%% BODY OF PAPER %%%%%%%%%%%%%%%%%%

\section{Introduction}

\label{sec:int}

At the long cooling era, the evolution of a neutron star (NS) surface
temperature crucially depends on the overall rate of neutrino emission out
of the star. The cooling dynamics below the superfluid transition
temperature is governed primarily by the superfluid component of nucleon
matter. The superfluidity of nucleons in NSs strongly suppresses most
mechanisms of neutrino emission operating in the non-superfluid nucleon
matter (the bremsstrahlung at nucleon collisions, modified Urca processes
etc. \cite{Friman1979,Tsuruta1986}) but simultaneously strongly reduces the
heat capacity and triggers the emission of neutrino pairs through neutral
weak currents caused by the nucleon Cooper pair breaking and formation (PBF)
processes in thermal equilibrium. Neutrino emission from Cooper pairs is
currently thought to be the dominant cooling mechanism of baryon matter, for
some ranges of the temperature and/or matter density. The total energy $%
\omega =\omega _{1}+\omega _{2}$ and momentum $\mathbf{k=k}_{1}+\mathbf{k}%
_{2}$ of an escaping (massless) neutrino pair form a time-like four-momentum 
$K=\left( \omega ,\mathbf{k}\right) $, so the process is kinematically
allowed only because of the existence of a superfluid energy gap $\Delta $,
that admits the nucleon transitions with $\omega >2\Delta $ and $k<\omega $.
(We use the Standard Model of weak interactions, the system of units $\hbar
=c=1$ and the Boltzmann constant $k_{B}=1$.)

The simplest case for baryon pairing corresponds to two particles correlated
in the $^{1}$S$_{0}$ state with the total spin $S=0$ and orbital momentum $%
L=0$. The neutrino emissivity due to the PBF processes in the spin-singlet
superfluid nucleon matter was first suggested and calculated by Flowers et
al. \cite{Flowers1976}. The result of this calculation was recovered later
by other authors \cite{Voskresensky1986,Voskresensky1987,Yakovlev1999}.
Similar mechanism for the neutrino energy losses due to spin-singlet pairing
of hyperons was suggested in \cite{Schaab1997,Schaab1998,Balberg1998}. More
than three decades these ideas was a key ingredient in numerical simulations
of NS evolution (e.g. \cite{Page1998,Yakovlev1998,Page2004}). However, after
such a long period, it was unexpectedly found that the PBF emission of
neutrino pairs is practically absent in a non-relativistic spin-singlet
superfluid liquid \cite{Leinson2006}. Later this result was confirmed in
other calculations \cite{Leinson2008,Kolomeitsev2008,Steiner2009}. (Note
also the controversial work \cite{Sedrakian2007}.)

The importance of the suppression of the PBF neutrino emission from the $%
^{1} $S$_{0}$ superfluid was first understood in \cite{Gupta2007} in
connection with the fact that the previous theory predicted a too rapid
cooling of the NS's crust, which dramatically contradicts the observed data
of superbursts \cite{Cumming2006}.

The $^{1}$S$_{0}$ neutron pairing in NS is essentially restricted to the
crust. As a result, in the NS evolution, effects of the suppression are
mostly observed during the thermal relaxation of the crust \cite%
{Lattimer1994,Page2009a,Page2009}. The significant revision of PBF neutrino
emission from this relatively thin layer does not change substantially the
total energy losses from the star. The most neutrino losses occur from the
NS core, which occupies more than 90\% of the star's volume and contains the
superfluid neutrons paired in the $^{3}$P$_{2}$ state with $S=1$, $L=1$ and $%
J=2$ \cite{Tamagaki1970,Takatsuka1972}.

In the commonly used version of the minimal cooling paradigm, the emission of $%
^{3} $P$_{2}$ pairing was reduced by only about 30\% due to the suppression
of the the vector channel of weak interactions \cite%
{Page2009,Ofengeim2015,Ofengeim2017}. This approach does not take into
account the anomalous axial-vector weak interactions, existing due to spin
fluctuations in the spin-triplet superfluid neutron matter \cite{Leinson2010}%
. Some simulations of the NS evolution accounting for the anomalous
contributions predict a raising of its surface temperature and argue that a
full exploration of this effect is necessary \cite{Han2017}. (Also see \cite%
{Shternin2015,Potekhin2015}).

A correct description of the efficiency of neutrino emission in the PBF
processes allows for a better understanding of observations \cite%
{Page2011,Shternin2011,Leinson2015}. This review is devoted to the current
state of this problem. Since the complete calculations have been published
repeatedly (e.g.\ \cite{Leinson2006,Leinson2010,Leinson2012}), I will
briefly sketch the main steps of the derivation, referring the reader to the
original papers for more detailed information.

\section{Preliminary notes}

\label{sec:prelim}

The low-energy Hamiltonian of the weak interaction may be described in a
point-like approximation. For interactions mediated by neutral weak
currents, it can be written as (e.g. \cite{Friman1979}) 
\begin{equation}
\mathcal{H}_{vac}=-\frac{G_{F}}{2\sqrt{2}}J_{B}^{\mu }l_{\mu }.  \label{L}
\end{equation}%
Here $G_{F}$ is the Fermi coupling constant, and the neutrino weak current
is given by $l_{\mu }=\bar{\nu}\gamma _{\mu }\left( 1-\gamma _{5}\right) \nu 
$, where $\gamma _{\mu }$ are Dirac matrices ($\mu =0,1,2,3$), $\gamma
_{5}=-i\gamma _{0}\gamma _{1}\gamma _{2}\gamma _{3}$. The neutral weak
current of the baryon, $J_{\mu }=C_{V}J_{\mu }^{V}-C_{A}J_{\mu }^{A}$,
represents the combination of the vector and axial-vector terms, $J_{\mu
}^{V}=\bar{\psi}\gamma _{\mu }\psi $ and $J_{\mu }^{A}=\bar{\psi}\gamma
_{\mu }\gamma _{5}\psi $, respectively. Here $\psi $ represents the baryon
field. The weak coupling constants $C_{V}$ and $C_{A}$ are determined by
quark composition of the baryons. For the reactions with neutrons, one has $%
C_{V}=1$ and $C_{A}=g_{A}$, while for those with protons, $C_{V}\simeq -0.08$
and $C_{A}=-g_{A}$, where $g_{A}\simeq 1.26$ is the axial--vector constant.
Notice that similar interaction Hamiltonian, but with other coupling constants, describes the neutrino weak interaction of hyperons in NS matter 
(e.g., \cite{Okun}).

In the non-relativistic nucleon system, the vector part of the weak current
can be approximated by its temporal component%
\begin{equation}
J_{0}^{V}=\psi ^{+}\hat{1}\psi ,  \label{jv}
\end{equation}%
where $\hat{1}=\delta _{\alpha \beta }$. Throughout the text, a hat means a $%
2\times 2$ matrix in spin space, $\alpha ,\beta =\uparrow ,\downarrow $. The
axial weak current is given dominantly by its space component 
\begin{equation}
\mathbf{J}^{A}=\psi ^{+}\boldsymbol{\hat{\sigma}}\psi ,  \label{ja}
\end{equation}%
where $\mathbf{\hat{\sigma}}=\left( \hat{\sigma}_{1},\hat{\sigma}_{2},\hat{%
\sigma}_{3}\right) $ are Pauli spin matrices.

It is important to notice that the vector weak current is conserved in the
standard theory. The conservation law implies that the transition matrix
element in the vector channel of the reaction obeys the relation%
\begin{equation}
\omega \left( J_{0}^{V}\right) _{fi}=\mathbf{k}\left( \mathbf{J}^{V}\right)
_{fi}.  \label{con}
\end{equation}%
The transferred momentum $\mathbf{k}$ enters into the medium response
function through the quasiparticle energy, which for $k\ll p_{F}$ in a
degenerate Fermi liquid takes the form $\xi _{\mathbf{p+k}}\simeq
v_{F}\left( p-p_{F}\right) +\mathbf{k}\mathbf{v}_{F}$. Thus, in the
absence of external fields, the momentum transfer $\mathbf{k}$ enters the
response function of the medium only in combination with the Fermi velocity,
which is small in the non-relativistic system, $v_{F}\ll 1$. Therefore, for
the PBF processes the relation $kv_{F}\ll \omega ,\Delta $ is always
satisfied. This allows one to evaluate the medium response function in the
long-wave limit $\mathbf{k}\rightarrow 0$. Together with the conservation law (\ref%
{con}) this immediately yields $\left( J_{0}^{V}\right) _{fi}=0$ for $\omega
>2\Delta $, which means that the neutrino pair emission through the vector
channel of weak interactions is strongly suppressed in the non-relativistic
system. This important fact was overlooked for a long time, since a direct
calculation shows that the matrix element $\left( J_{0}^{V}\right) _{fi}$
for the recombination of two Bogolons into the condensate does not vanish,
which erroneously leads to a large neutrino emissivity through the vector
channel.

First calculations of the PBF neutrino energy losses were performed using a
vacuum-type weak interactions assuming that the medium effects can be taken
into account by introducing effective masses of participating quasiparticles 
\cite{Flowers1976,Yakovlev1999}. This resulted to a substantial overestimate
of the PBF neutrino energy losses from the superfluid core and inner crust
of NSs. Only three decades later it has been understood that the calculation
of neutrino radiation from a superfluid Fermi liquid requires a more
delicate approach.

Within the Nambu-Gor'kov formalism the effective vertex of nucleon
interactions with an external neutrino field represents a $2\times 2$ matrix
in the particle-hole space. This matrix is diagonal for nucleons in the
normal Fermi liquid but it gets the off-diagonal entries in superfluid
systems \cite{Bogoliubov,Nambu,Larkin,Leggett}. The diagonal elements
represent the ordinary (dressed) vertices of the field interaction with
quasiparticles and holes, respectively, while the off-diagonal elements of
the matrix represent the effective vertices for a virtual breaking and
formation of Cooper pairs in the external field. In other words, the
off-diagonal components of the vertex matrix describe a coupling of the
external field with fluctuations of the order parameter in the superfluid
Fermi liquid. These so-called "anomalous weak interactions" should be
necessarily taken into account when calculating the neutrino energy losses
from superfluid cores of NSs.

In particular, the anomalous weak interactions are crucial for the neutrino
emission caused by the PBF processes. For example, in non-relativistic
systems, the ordinary and anomalous contributions into the matrix element of
the weak vector transition current mutually cancel in the long-wave limit,
leading to a strong suppression of the PBF neutrino emission \cite%
{Leinson2006}. The more accurate calculation \cite{Leinson2008,Steiner2009}
has shown, the neutrino-pair emission owing to the density fluctuations is
suppressed proportionally to $v_{F}^{4}$. This reflects the well known fact
that the dipole radiation is not possible in the vector channel in the
collision of two identical particles. Thus, exactly due to the anomalous
contributions the PBF neutrino emission in the vector channel of weak
interactions is practically absent.

In the case of $^{1}$S$_{0}$ pairing this has far-reaching consequences. The
total spin $\mathbf{S}=0$ of the non-relativistic Cooper pair is conserved.
Therefore the neutrino emission through the axial-vector channel of weak
interactions could arise only due to small relativistic effects and is
proportional to $v_{F}^{2}$ \cite{Flowers1976,Kolomeitsev2008}. Thus the PBF
neutrino energy losses due to singlet-state pairing of baryons can, in
practice, be neglected in simulations of NS cooling. This makes unimportant
the neutrino radiation from $^{1}$S$_{0}$ pairing of protons or hyperons.

The minimal cooling paradigm \cite{Page2009} suggests that, below the
critical temperature for a triplet pairing of neutrons, the dominant
neutrino energy losses occur from the superfluid neutron liquid in the inner
core of a NS. It is commonly believed \cite%
{Tamagaki1970,Hoffberg1970,Takatsuka1972,Baldo1992,Elgaroy1996} that, in
this case, the $^{3}$P$_{2}$ pairing (with a small admixture of $^{3}$F$_{2}$
state) takes place with a preferred magnetic quantum number $M_{J}=0$. Since
the spin of a Cooper pair in the $^{3}$P$_{2}$ state is $S=1$ the spin
fluctuations are possible and the PBF neutrino energy losses from the
neutron superfluid occur through the axial channel of weak interactions.

The pairing interaction, in the most attractive $^{3}$P$_{2}$ channel, can
be written as \cite{Tamagaki1970} 
\begin{equation}
\Gamma _{\alpha \beta ,\gamma \delta }\left( \mathbf{p,p}^{\prime }\right) =%
\frac{\pi ^{2}}{p_{F}m^{\ast }}V\left( p,p^{\prime }\right) \left[ \mathbf{%
\bar{b}}\left( \mathbf{n}\right) \mathbf{\hat{\sigma}}i\hat{\sigma}_{2}%
\right] _{\alpha \beta }\left[ i\hat{\sigma}_{2}\mathbf{\hat{\sigma}\bar{b}}(%
\mathbf{n}^{\prime })\right] _{\gamma \delta }~,  \label{ppint}
\end{equation}%
where $V\left( p,p^{\prime }\right) $ is the corresponding interaction
amplitude; $p_{F}$ and $m^{\ast }=p_{F}/v_{F}$ are the Fermi momentum and
the neutron effective mass, respectively, so that $p_{F}m^{\ast }/\pi ^{2}$
is the density of states near the Fermi surface. The angular dependence of
the interaction is represented by Cartesian components of the unit vector $%
\mathbf{n=p}/p$ which involves the polar angles on the Fermi surface, 
\begin{equation}
n_{1}=\sin \theta \ \cos \varphi ,\ \ \ n_{2}=\sin \theta \ \sin \varphi ,\
\ \ n_{3}=\cos \theta .  \label{n}
\end{equation}%
Further, $\mathbf{\bar{b}}\left( \mathbf{n}\right) $ is a real vector in the
spin space, normalizable by condition 
\begin{equation}
\left\langle \bar{b}^{2}\left( \mathbf{n}\right) \right\rangle =1~.
\label{Norm}
\end{equation}%
Hereafter we use the angle brackets to denote angle averages, 
\begin{equation}
\left\langle ...\right\rangle \equiv \frac{1}{4\pi }\int d\mathbf{n}\cdot
\cdot \cdot =\frac{1}{2}\int_{-1}^{1}dn_{3}\int_{0}^{2\pi }\frac{d\varphi }{%
2\pi }\cdot \cdot \cdot .  \label{av}
\end{equation}

For spin-triplet pairing, the order parameter $\hat{D}\equiv D_{\alpha \beta
}\left( \mathbf{n}\right) $ is a symmetric matrix in the spin space, which
near the Fermi surface can be written as (see e.g. \cite{Ketterson}) 
\begin{equation*}
\hat{D}\left( \mathbf{n},T\right) =\Delta \,\mathbf{\bar{b}\hat{\sigma}}i%
\hat{\sigma}_{2},
\end{equation*}%
where the temperature-dependent gap amplitude $\Delta \left( T\right) $ is a
real constant.

The vector $\mathbf{\bar{b}}$ defines the angle anisotropy of energy gap
which depends on the phase state of the superfluid condensate. In general,
this vector can be written in the form $\bar{b}_{i}=\bar{A}_{ij}n_{j}$,
where $\bar{A}_{ij}$ is a $3\times 3$ matrix. In the case of a unitary $^{3}$%
P$_{2}$ condensate the matrix $\bar{A}_{ij}$ must be a real symmetric
traceless tensor. It may be specified by giving the orientation of its
principal axes and its two independent diagonal elements in its
principal-axis coordinate system. Within the preferred coordinate system,
the ground state with $M_{J}=0$ is described by the matrix 
\begin{equation}
\bar{A}_{ij}=\frac{1}{\sqrt{2}}\mathsf{diag}\left( -1,-1,2\right)  \label{A}
\end{equation}%
and $\bar{b}^{2}(\mathbf{n})=1/2\left( 1+3\cos ^{2}\theta \right) $.

\section{General approach to neutrino energy losses}

\label{sec:appr}

Thermal fluctuations of the neutral weak currents in nucleon matter are
closely related to the imaginary, dissipative part of the response function
of the medium onto external the neutrino field. According to the
fluctuation-dissipation theorem, the total energy loss per unit volume and
time caused by thermal fluctuations of the neutral weak current in the
nucleon matter is given by the following formula 
\begin{equation}
Q=\frac{G_{F}^{2}}{8}\sum_{\nu }\int \;\omega \frac{2\mathrm{Im}\Pi _{\mu
\nu }\left( \omega ,\mathbf{k}\right) \mathrm{Tr}\left( l^{\mu }l^{\nu \ast
}\right) }{1-\exp \left( \omega /T\right) }\frac{d^{3}k_{1}}{2\omega
_{1}(2\pi )^{3}}\frac{d^{3}k_{2}}{2\omega _{2}(2\pi )^{3}},  \label{Qnu}
\end{equation}%
where $\mathrm{Im}\Pi _{\mu \nu }$ is the imaginary part of the retarded
weak polarization tensor. The integration goes over the phase volume of
neutrinos and antineutrinos of total energy $\omega =\omega _{1}+\omega _{2}$
and total momentum $\mathbf{k}\mathbf{=k}_{1}+\mathbf{k}_{2}$. The symbol $%
\sum_{\nu }$\ \ indicates a summation over three neutrino flavors. The
factor $\left[ 1-\exp \left( \omega /T\right) \right] ^{-1}$ occurs as a
result of averaging over the Gibbs distribution, which must be performed at
finite ambient temperatures.

By inserting $\int d^{4}K\delta ^{\left( 4\right) }\left(
K-K_{1}-K_{2}\right) =1$ in this equation, and making use of the Lenard's
integral \ 
\begin{equation}
\int \frac{d^{3}k_{1}}{2\omega _{1}}\frac{d^{3}k_{2}}{2\omega _{2}}\delta
^{\left( 4\right) }\left( K-K_{1}-K_{2}\right) \mathrm{Tr}\left( l^{\mu
}l^{\nu \ast }\right) =\frac{4\pi }{3}\left( K^{\mu }K^{\nu }-K^{2}g^{\mu
\nu }\right) \Theta \left( K^{2}\right) \Theta \left( \omega \right) ,
\label{LI}
\end{equation}%
where $K_{1}=\left( \omega _{1},\mathbf{k}_{1}\right) $, $K_{2}=\left(
\omega _{2},\mathbf{k}_{2}\right) $,$~\Theta (x)$ is the Heaviside step
function, and $g^{\mu \nu }=\mathsf{diag}(1,-1,-1,-1)$ is the signature
tensor, we can write 
\begin{equation}
Q=\frac{G_{F}^{2}\mathcal{N}_{\nu }}{192\pi ^{5}}\int_{0}^{\infty }d\omega
\int d^{3}k\frac{\omega \Theta \left( \omega -k\right) }{1-\exp \left(
\omega /T\right) }\mathrm{Im}\Pi _{\mu \nu }\left( \omega ,\mathbf{k}\right)
\left( K^{\mu }K^{\nu }-K^{2}g^{\mu \nu }\right) ~,  \label{QQQ}
\end{equation}%
where $\mathcal{N}_{\nu }=3$ is the number of neutrino flavors.

In general, the weak polarization tensor of the medium is a sum of the
vector-vector, axial-axial, and mixed terms. The mixed vector-axial
polarization has to be an antisymmetric tensor, and its contraction in Eq. (%
\ref{QQQ}) with the symmetric tensor $K^{\mu }K^{\nu }-K^{2}g^{\mu \nu }$
vanishes. Thus only the pure-vector and pure-axial polarizations should be
taken into account. We then obtain 
\begin{equation}
\mathrm{Im}\Pi _{\mu \nu }=C_{V}^{2}\mathrm{Im}\Pi _{\mu \nu }^{V}+C_{A}^{2}%
\mathrm{Im}\Pi _{\mu \nu }^{A},  \label{imP}
\end{equation}%
where $C_{V}$ and $C_{A}$ are vector and axial-vector weak coupling
constants of a neutron, respectively.

\section{Weak interactions in superfluid Fermi liquids}

\label{sec:weak}

\begin{figure}[h]
\includegraphics{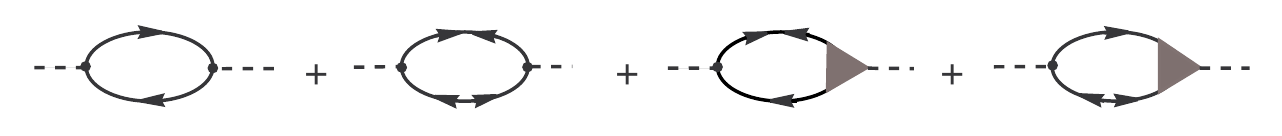}
\caption{Graphs for the polarization tensor. }
\label{fig1}
\end{figure}

Physically, the polarization tensor represents a correction to the Z-boson
self-energy in the medium. Making use of the adopted graphical notation for
the ordinary and anomalous propagators, $\hat{G}=\parbox{1cm}{%
\includegraphics[width=1cm]{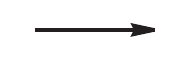}}$, $\hat{G}^{-}(p)=\parbox{1cm}{%
\includegraphics[width=1cm,angle=180]{gn.pdf}}$, $\hat{F}^{(1)}=%
\parbox{1cm}{\includegraphics[width=1cm]{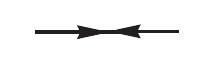}}$\thinspace , and $\hat{F}%
^{(2)}=\parbox{1cm}{\includegraphics[width=1cm]{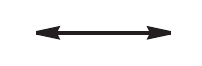}}$\thinspace , one
can represent the polarization function in each of the channels as the sum
of graphs depicted in Fig. \ref{fig1}. 

As can be seen, the field interaction with superfluid fermions should be
described with the aid of four effective three-point vertices. There are two
usual effective vertices (shown by dots) corresponding to the creation of a
particle and a hole by the Z-field. Let us denote them as $\hat{\tau}\left( 
\mathbf{n;}\omega ,\mathbf{k}\right) $ and $\hat{\tau}^{-}\left( \mathbf{n;}%
\omega ,\mathbf{k}\right) \equiv \hat{\tau}^{T}\left( -\mathbf{n;}\omega ,%
\mathbf{k}\right) $, respectively. We omit the Dirac indices in these
symbolic notations. In reality, according to Eqs. (\ref{jv}) and (\ref{ja}),
the non-relativistic ordinary vector vertex is represented by its temporal
component, i.e. it is a scalar matrix in spin space. The ordinary
axial-vector vertices of a particle and a hole are represented by
space-vectors which components consist of spin matrices.

Two more vertices, represented by triangles, correspond to the creation of
two particles or two holes. These so-called "anomalous" vertices appear
because the pairing interaction among quasi-particles is to be incorporated
in the coupling vertex up to the same degree of approximation as in the
self-energy of a quasiparticle \cite{Bogoliubov,Nambu}. This means that the
anomalous effective vertices are given by infinite sums of diagrams with
allowance for pair interaction in the ladder approximation, in the same way
as in the gap equations.

\begin{figure}[h]
\resizebox{0.7\hsize}{!}{\includegraphics{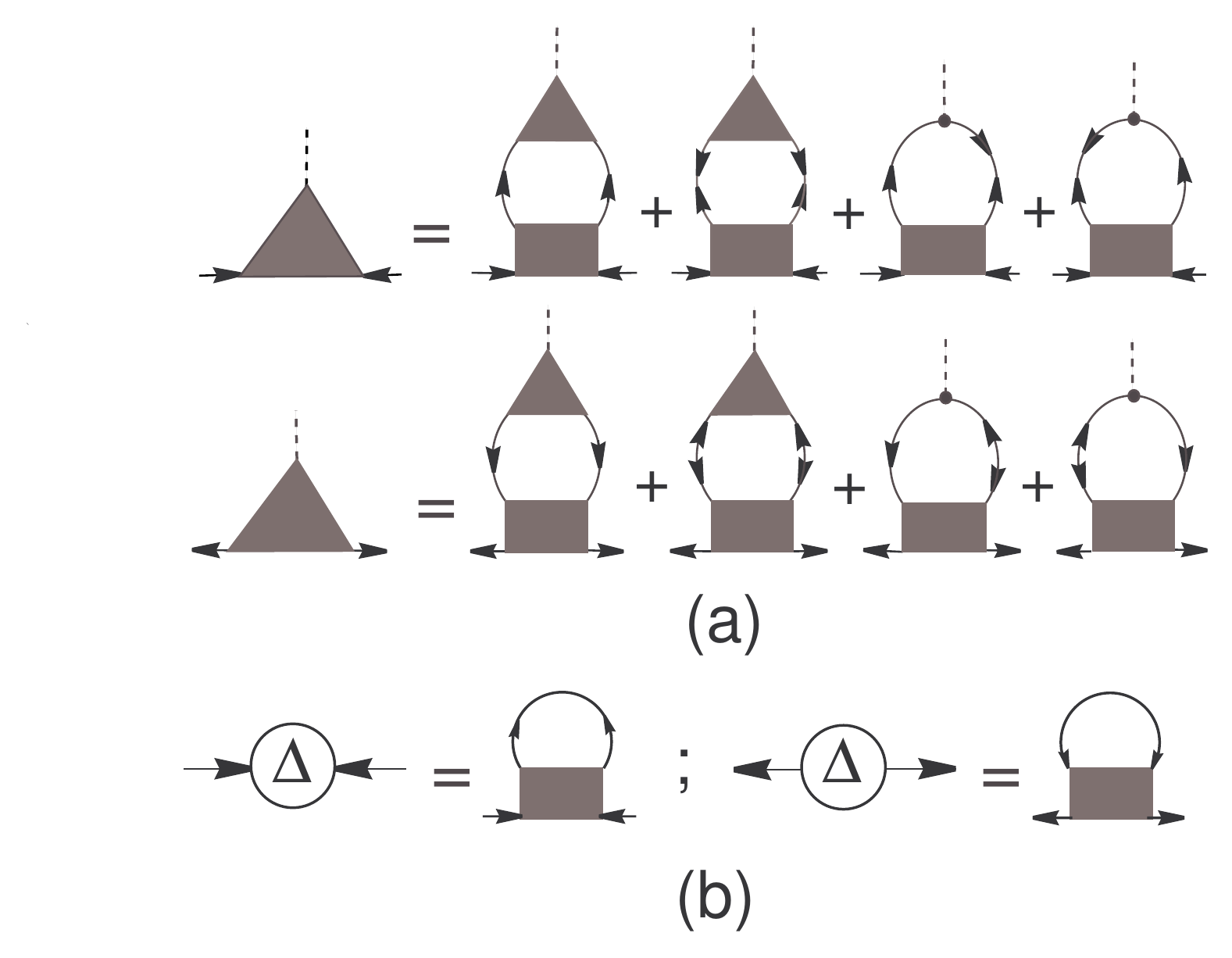}}
\caption{Dyson's equations for the anomalous vertices (a), and the gap
equations (b). Shaded rectangles represent the pairing interaction; }
\label{fig2}
\end{figure}

Given by the sum of ladder-type diagrams \cite{Larkin}, the anomalous
vertices are to satisfy the Dyson's equations symbolically depicted by
graphs in Fig. \ref{fig2}a. 
In these graphs, the rectangles denote pairing interaction, which in the
channel of two quasiparticles is given by Eq. (\ref{ppint}). The vertex
equations are to be supplemented by the gap equation shown graphically in
Fig. \ref{fig2}b. This equation, whose solution is assumed known, serves to
eliminate the amplitude of the pair interaction from the vertex equations
near the Fermi surface. The standard gap equation involves integrations over
the regions far from the Fermi surface. This integration can be eliminated
by means of the renormalization of the pairing interaction, as suggested in
Ref. \cite{Leggett}. Details of this calculation can be found in \cite%
{Leinson2012}.

The analytic form of the quasiparticle propagators in the momentum
representation can be written as 
\begin{equation}
\hat{G}=G\left( \varepsilon _{s},\mathbf{p}\right) \hat{1},~\ \ \hat{F}%
^{\left( 1\right) }=F\left( \varepsilon _{s},\mathbf{p}\right) \left( 
\mathbf{\bar{b}\hat{\sigma}}\right) i\hat{\sigma}_{2},  \label{eqpr}
\end{equation}%
\begin{equation}
\hat{F}^{\left( 2\right) }=\hat{F}^{\left( 1\right) \dagger }\left(
-\varepsilon _{s},-\mathbf{p}\right) =i\hat{\sigma}_{2}\left( \mathbf{\bar{b}%
\hat{\sigma}}\right) F\left( -\varepsilon _{s},-\mathbf{p}\right) .
\label{F2}
\end{equation}%
Making use of the Matsubara calculation technique we define the scalar part
of the Green functions%
\begin{equation}
G\left( \varepsilon _{s},\mathbf{p}\right) =\frac{-i\varepsilon _{s}-\xi _{p}%
}{\varepsilon _{s}^{2}+E_{\mathbf{p}}^{2}},~\ \ F\left( \varepsilon _{s},%
\mathbf{p}\right) =\frac{\Delta }{\varepsilon _{s}^{2}+E_{\mathbf{p}}^{2}}.
\label{GF}
\end{equation}%
Here $\varepsilon _{s}=\left( 2s+1\right) \pi T$ with $s=0,\pm 1,\pm 2,...$
be the fermionic Matsubara frequency which depends on the temperature $T$,
and%
\begin{equation}
E=\sqrt{\xi ^{2}+\Delta _{\mathbf{n}}^{2}}~  \label{E}
\end{equation}%
stands for the Bogolon energy. The angle-dependent energy gap is given by $%
\Delta _{\mathbf{n}}^{2}\equiv \Delta ^{2}\bar{b}^{2}\left( \mathbf{n}%
\right) $.

It should be noted that, by virtue of Eq. (\ref{Norm}), the amplitude $%
\Delta \left( T\right) $ is chosen as to represent the energy gap averaged
over the Fermi surface. Thus determined, the energy gap gives a general
measure of the pairing correction to the energy of the ground state in the
preferred state.

In general, the ordinary vertices in the Dyson equations should be dressed
owing to residual Fermi-liquid interactions. We neglect this effect, and
account for the residual interactions by means of the effective nucleon mass
only. In this case the ordinary vertices are as defined in Eqs. (\ref{jv}), (%
\ref{ja}). Namely, the non-relativistic ordinary vector vertex is
represented by its temporal component 
\begin{equation}
\hat{\tau}_{V}=\hat{\tau}_{V}^{-}=\hat{1}.  \label{tauV}
\end{equation}%
The ordinary axial-vector vertices of a particle and a hole are to be taken
as 
\begin{equation}
\hat{\tau}_{A}=\mathbf{\hat{\sigma},~}\hat{\tau}_{A}^{-}=\mathbf{\hat{\sigma}%
}^{T},  \label{tauA}
\end{equation}%
where the upperscript "$T$" transposes the matrix.

In the case of pairing in the channel with spin, orbital and total angular
momenta, $S=1,L=1,J=2$, respectively, one can search for the anomalous
vertices near the Fermi surface in the form of expansions over the
eigenfunctions of the total angular momentum $(J,M)$ with $J=2$ and $M=0,\pm
1,\pm 2$. For our calculations it is convenient to use vector notation which
involves a set of mutually orthogonal complex\ vectors $\mathbf{b}_{M}\left( 
\mathbf{n}\right) $ in spin space which generate standard spin-angle
matrices according to 
\begin{equation}
\frac{1}{\sqrt{8\pi }}\mathbf{b}_{M}(\mathbf{n})\mathbf{\hat{\sigma}}i\hat{%
\sigma}_{2}\equiv \sum_{M_{S}+M_{L}=M}\left( \frac{1}{2}\frac{1}{2}\alpha
\beta |1M_{S}\right) \left( 11M_{S}M_{L}|2M\right) Y_{1,M_{L}}\left( \mathbf{%
n}\right) ,  \label{bm}
\end{equation}%
where $\alpha ,\beta =\uparrow ,\downarrow $ denote spin projections.

These vectors are of the form 
\begin{align}
\mathbf{b}_{0}& =\sqrt{1/2}\left( -n_{1},-n_{2},2n_{3}\right) ,\mathbf{b}%
_{1}=-\sqrt{3/4}\left( n_{3},in_{3},n_{1}+in_{2}\right) ,  \notag \\
\mathbf{b}_{2}& =\sqrt{3/4}\left( n_{1}+in_{2},in_{1}-n_{2},0\right) ,%
\mathbf{b}_{-M}=\left( -\right) ^{M}\mathbf{b}_{M}^{\ast }.  \label{b012}
\end{align}%
These are normalized by the condition%
\begin{equation}
\left\langle \mathbf{b}_{M^{\prime }}^{\ast }\mathbf{b}_{M}\right\rangle
=\delta _{M^{\prime }M}.  \label{lmnorm}
\end{equation}

Generally speaking, the anomalous vertices are functions of the transferred
energy and momentum $(\omega ,\mathbf{k})$ and the direction 
$\mathbf{n}$ of the quasiparticle momentum. As was mentioned in Introduction, 
it is sufficient to evaluate the medium response function in the limit 
$\mathbf{k}\rightarrow 0$. Then the non-relativistic anomalous vector vertex 
can be expanded in the eigenfunctions of the total angular momentum 
$J=2$ in the form 
\begin{equation}
\mathcal{\hat{T}}^{\left( 1\right) }=\sum_{M}\mathcal{B}_{M}^{\left(
1\right) }\left( \omega \right) \mathbf{b}_{M}\mathbf{\hat{\sigma}}i\hat{%
\sigma}_{2},  \label{Tm1}
\end{equation}%
\begin{equation}
\mathcal{\hat{T}}^{\left( 2\right) }=\sum_{M}\mathcal{B}_{M}^{\left(
2\right) }\left( \omega \right) i\hat{\sigma}_{2}\mathbf{\hat{\sigma}b}_{M}.
\label{Tm2}
\end{equation}%
Accordingly, the anomalous axial-vector vertices can be represented in the
form%
\begin{equation}
\mathbf{\hat{T}}^{\left( 1\right) }=\sum_{M}\mathbf{B}_{M}^{\left( 1\right)
}\left( \omega \right) \mathbf{b}_{M}\mathbf{\hat{\sigma}}i\hat{\sigma}_{2},
\label{Ta1}
\end{equation}%
\begin{equation}
\mathbf{\hat{T}}^{\left( 2\right) }=\sum_{M}\mathbf{B}_{M}^{\left( 2\right)
}\left( \omega \right) i\hat{\sigma}_{2}\mathbf{\hat{\sigma}b}_{M}.
\label{Ta2}
\end{equation}

Making use of these general forms in the Dyson equations together with the
corresponding ordinary vertices, after tedious computations, one can get 
\cite{Leinson2012} in the vector channel%
\begin{equation}
\mathcal{B}_{M}^{\left( 1\right) }=-\mathcal{B}_{M}^{\left( 2\right) }\equiv 
\mathcal{B}_{M},  \label{BmV}
\end{equation}%
where $\mathcal{B}_{M}$ obeys the equation 
\begin{align}
& \sum_{M^{\prime }}2\left\langle \left[ \left( \Omega ^{2}-\bar{b}%
^{2}\right) \mathbf{b}_{M}^{\ast }\mathbf{b}_{M^{\prime }}+\left( \mathbf{b}%
_{M}^{\ast }\mathbf{\bar{b}}\right) \left( \mathbf{b}_{M^{\prime }}\mathbf{%
\bar{b}}\right) \right] \mathcal{I}_{0}\right\rangle \mathcal{B}_{M^{\prime
}}  \notag \\
& +\left\langle \left( b_{M}^{2}-\bar{b}^{2}\right) A\right\rangle \mathcal{B%
}_{M}=2\Omega \left\langle \left( \mathbf{b}_{M}^{\ast }\mathbf{\bar{b}}%
\right) \mathcal{I}_{0}\right\rangle ,  \label{Bm}
\end{align}%
In the axial-vector channel one finds%
\begin{equation}
\mathbf{B}_{M}^{\left( 1\right) }=\mathbf{B}_{M}^{\left( 2\right) }\equiv 
\mathbf{B}_{M}  \label{B12B}
\end{equation}%
with $\mathbf{B}_{M}$ satisfying the equation 
\begin{align}
& \sum_{M^{\prime }}~2\left\langle \left[ \Omega ^{2}\mathbf{b}_{M}^{\ast }%
\mathbf{b}_{M^{\prime }}-\left( \mathbf{b}_{M}^{\ast }\mathbf{\bar{b}}%
\right) \left( \mathbf{b}_{M^{\prime }}\mathbf{\bar{b}}\right) \right] 
\mathcal{I}_{0}\right\rangle \mathbf{B}_{M^{\prime }}  \notag \\
& +\left\langle \left( b_{M}^{2}-\bar{b}^{2}\right) A\right\rangle \mathbf{B}%
_{M}=-2i\Omega \left\langle \left( \mathbf{b}_{M}^{\ast }\mathbf{\times \bar{%
b}}\right) \mathcal{I}_{0}\right\rangle ,  \label{Bp}
\end{align}%
In the above expressions, the following notation is used:%
\begin{equation}
\Omega =\frac{\omega }{2\Delta },  \label{Eta}
\end{equation}%
the functions $\mathcal{I}_{0}\left( \omega ,\mathbf{n},T\right) $ and $%
A\left( \mathbf{n},T\right) $ are given by 
\begin{equation}
\mathcal{I}_{0}\left( \omega \mathbf{,n}\right) =\int_{-\infty }^{\infty }%
\frac{d\xi }{E}\frac{\Delta ^{2}}{4E^{2}-\left( \omega +i0\right) ^{2}}\tanh 
\frac{E}{2T}~,  \label{FFq0}
\end{equation}%
\begin{equation}
A\left( \mathbf{n}\right) \equiv \int_{-\infty }^{\infty }d\xi \left( \frac{1%
}{2E}\tanh \frac{E}{2T}-\frac{1}{2\xi }\tanh \frac{\xi }{2T}\right) .
\label{Aex}
\end{equation}

From Eqs. (\ref{Bm}) and (\ref{Bp}) it is seen that an accurate calculation
of the anisotropic anomalous vertices at arbitrary temperatures apparently
requires numerical computations. It would be desirable, however, to get
reasonable analytic expressions for the anomalous vertices, which can be
applied to a calculation of the neutrino energy losses. To proceed, let us
notice that the anisotropy of the functions $\mathcal{I}_{0}\left( \omega 
\mathbf{,n}\right) $ and $A\left( \mathbf{n}\right) $ is due to the
dependence of the energy of the Bogolons (\ref{E}) on the direction of the
momentum relative to the quantization axis. In a uniform system without
external fields and at absolute zero, the orientation of the quantization
axis is arbitrary. For equilibrium at a non-zero temperature this leads to
the formation of a loose domain structure \cite{br60}, where each
microscopic domain has a randomly oriented preferred axis. This fact is
normally used in order to simplify the calculations by replacing the
angle-dependent energy gap with some effective isotropic value (see, e.g. 
\cite{bhy01,gh05}).

Making use of this trick we replace the angle-dependent energy gap $\Delta _{%
\mathbf{n}}^{2}\equiv \Delta ^{2}\bar{b}^{2}\left( \mathbf{n}\right) $ in
the Bogolons energy by its average value $\left\langle \Delta ^{2}\bar{b}%
^{2}\left( \mathbf{n}\right) \right\rangle =\Delta ^{2}$, in accordance with
Eq. (\ref{Norm}). Then the functions $\mathcal{I}_{0}$ and $A$ can be moved
out the integrals over the solid angle in Eqs. (\ref{Bm}) and (\ref{Bp}).
Using further the axial symmetry of the order parameter, Eq. (\ref{lmnorm})
and the fact that 
\begin{equation}
\left\langle \left( \mathbf{b}_{M}^{\ast }\mathbf{b}_{M}-\bar{b}^{2}\right)
\right\rangle =0~  \label{Am}
\end{equation}%
we get for the vector channel the equation 
\begin{equation}
\left( \Omega ^{2}-\left\langle \bar{b}^{2}\mathbf{b}_{M}^{\ast }\mathbf{b}%
_{M}\right\rangle \right) \mathcal{B}_{M}+\sum_{M^{\prime }}\left\langle
\left( \mathbf{b}_{M}^{\ast }\mathbf{\bar{b}}\right) \left( \mathbf{b}%
_{M^{\prime }}\mathbf{\bar{b}}\right) \right\rangle \mathcal{B}_{M^{\prime
}}=\Omega \left\langle \mathbf{b}_{M}^{\ast }\mathbf{\bar{b}}\right\rangle .
\label{vB}
\end{equation}%
In the axial channel we obtain the equation 
\begin{equation}
\Omega ^{2}\mathbf{B}_{M}-\sum_{M^{\prime }}\left\langle \left( \mathbf{b}%
_{M}^{\ast }\mathbf{\bar{b}}\right) \left( \mathbf{b}_{M^{\prime }}\mathbf{%
\bar{b}}\right) \right\rangle \mathbf{B}_{M^{\prime }}=-i\Omega \left\langle
\left( \mathbf{b}_{M}^{\ast }\mathbf{\times \bar{b}}\right) \right\rangle .
\label{aB}
\end{equation}%
The specific form of solutions to Eqs. (\ref{vB}) and (\ref{aB}) depends on
the phase state of the condensate.

An inspection of Eqs (\ref{A}) and (\ref{b012}) allows one to conclude that
for the ground state with $M_{J}=0$%
\begin{equation}
\mathbf{\bar{b}}_{M=0}\mathbf{=b}_{0}.  \label{bbar}
\end{equation}%
In this case we get $\left\langle \mathbf{b}_{M}^{\ast }\mathbf{\bar{b}}%
\right\rangle =\delta _{M,0}$, and the only non-vanishing values of $%
\left\langle \mathbf{b}_{M}^{\ast }\mathbf{\times \bar{b}}\right\rangle $
correspond to $M=\pm 1$. Simple calculations give%
\begin{equation}
\mathcal{B}_{M}=\frac{2\Delta }{\omega }\delta _{M,0}  \label{BV0}
\end{equation}%
and%
\begin{equation}
~\mathbf{B}_{1}=\sqrt{\frac{3}{2}}\frac{\Delta \omega }{\omega ^{2}-\Delta
^{2}/5}\mathbf{e}^{\ast },~\mathbf{B}_{-1}=\sqrt{\frac{3}{2}}\frac{\Delta
\omega }{\omega ^{2}-\Delta ^{2}/5}\mathbf{e}  \label{BA0}
\end{equation}%
where 
\begin{equation}
\mathbf{e}=\left( 1,i,0\right) .  \label{e}
\end{equation}

Substituting the obtained expressions to Eqs. (\ref{Tm1}) - (\ref{Ta2}) we
get the anomalous vertices which, together with the ordinary vertices (\ref%
{tauV}) and (\ref{tauA}), can be used to calculate the weak polarization
tensor of the medium. We now turn to a calculation of the corresponding
correlation functions separately in the vector and axial channel of weak
interactions.

\section{Correlation functions of weak currents}

\label{sec:corr}

\subsection{Vector channel}

Following to the graphs of Fig. \ref{fig1} the vector-vector part of the
polarization tensor, $\Pi _{\mu \nu }^{V}=\delta _{\mu 0}\delta _{\nu 0}\Pi
_{00}^{V}$, is given by analytic continuation of the following Matsubara
sums to the upper half-plane of the complex variable $\omega $:%
\begin{equation}
\Pi _{00}^{V}\left( \omega \right) =T\sum_{\mathbf{p,}\varepsilon _{s}}%
\mathrm{Tr}\left( \hat{G}_{+}\hat{G}+\hat{F}_{+}^{\left( 1\right) }\hat{F}%
^{\left( 2\right) }+\hat{F}_{+}^{\left( 1\right) }\mathcal{\hat{T}}^{\left(
1\right) }\hat{G}+\hat{G}_{+}\mathcal{\hat{T}}^{\left( 2\right) }\hat{F}%
^{\left( 2\right) }\right) .  \label{PiV}
\end{equation}%
We use the notations $\hat{G}_{+}=\hat{G}\left( \varepsilon _{s}+\omega _{n},%
\mathbf{p}\right) $, $\hat{F}_{+}^{\left( 1\right) }=\hat{F}^{\left(
1\right) }\left( \varepsilon _{s}+\omega _{n},\mathbf{p}\right) $, where $%
\omega _{n}=2i\pi Tn$ with $n=0,\pm 1,\pm 2...$ is a bosonic Matsubara
frequency.

The two first terms in the right of Eq. (\ref{PiV}) describe the medium
polarization without anomalous contributions. The \ long-wave limit of this
ordinary contribution in the vector channel can be found in the form 
\begin{equation*}
\left( \Pi _{00}^{V}\right) _{\mathsf{ordin}}\simeq -4\frac{p_{F}m^{\ast }}{%
\pi ^{2}}\left\langle \bar{b}^{2}\left( \mathbf{n}\right) \mathcal{I}%
_{0}\left( \mathbf{n,}\omega \right) \right\rangle
\end{equation*}%
Evidently this expression does not satisfy the condition of current
conservation $\omega \Pi _{00}^{V}=k_{i}\Pi _{i0}^{V}$, which in the
long-wave limit $\mathbf{k}\rightarrow 0$ requires $\Pi _{00}^{V}\left( \omega
>0\right) =0$.

The last two terms in Eq. (\ref{PiV}), with the vertices indicated in Eqs. (%
\ref{Tm1}), (\ref{Tm2}), represent the anomalous contributions. According to
Eqs. (\ref{BmV}) and (\ref{BV0}) the anomalous vector vertices can be
written as 
\begin{equation*}
\mathcal{\hat{T}}^{\left( 1\right) }=\frac{2\Delta }{\omega }\mathbf{\bar{b}%
\hat{\sigma}}i\hat{\sigma}_{2},
\end{equation*}%
and%
\begin{equation*}
\mathcal{\hat{T}}^{\left( 2\right) }=-\frac{2\Delta }{\omega }i\hat{\sigma}%
_{2}\mathbf{\hat{\sigma}\bar{b}}.
\end{equation*}%
Straightforward calculations give in the long-wave limit%
\begin{equation*}
\left( \Pi _{00}^{V}\right) _{\mathsf{anom}}\simeq 4\frac{p_{F}m^{\ast }}{%
\pi ^{2}}\left\langle \bar{b}^{2}\left( \mathbf{n}\right) \mathcal{I}%
_{0}\left( \mathbf{n,}\omega \right) \right\rangle .
\end{equation*}%
We finally find%
\begin{equation}
\Pi _{00}^{V}\left( \omega ,\mathbf{0}\right) =\left( \Pi _{00}^{V}\right) _{%
\mathsf{ordin}}+\left( \Pi _{00}^{V}\right) _{\mathsf{anom}}=0,  \label{PI00}
\end{equation}%
as is required by the current conservation condition. This proves explicitly
that the neutrino emissivity via the vector channel, as initially obtained
in \cite{Yakovlev1999}, is a subject of inconsistency.

\subsection{Axial channel}

In the axial channel, the ordinary vertices (\ref{tauA}) and anomalous
vertices (\ref{T1f}), (\ref{T2f}) consist of only space components, and thus 
$\Pi _{\mu \nu }^{A}\simeq \delta _{\mu i}\delta _{\nu j}C_{A}^{2}\Pi
_{ij}^{A}$, where $\Pi _{ij}^{A}$ is to be found as the analytic
continuation of the following Matsubara sums:%
\begin{eqnarray}
\Pi _{ij}^{A}\left( \omega \right) &=&T\sum_{\mathbf{p,}\varepsilon _{s}}%
\mathrm{Tr}\left( \hat{\sigma}_{i}\hat{G}_{+}\hat{\sigma}_{j}\hat{G}+\hat{%
\sigma}_{i}\hat{F}_{+}^{\left( 1\right) }\hat{\sigma}_{j}^{-}\hat{F}^{\left(
2\right) }\right)  \notag \\
&&+T\sum_{\mathbf{p,}\varepsilon _{s}}\mathrm{Tr}\left( \hat{\sigma}_{i}\hat{%
F}_{+}^{\left( 1\right) }\hat{T}_{j}^{\left( 1\right) }\hat{G}+\hat{\sigma}%
_{i}\hat{G}_{+}\hat{T}_{j}^{\left( 2\right) }\hat{F}^{\left( 2\right)
}\right) .  \label{PiA}
\end{eqnarray}%
Here the first line represents the ordinary contribution and the second line
is the contribution of the anomalous interactions. The ordinary contribution
can be evaluated in the form%
\begin{equation}
\left( \Pi _{ij}^{A}\right) _{\mathsf{ordin}}=-4\frac{p_{F}m^{\ast }}{\pi
^{2}}\left\langle \left( \bar{b}^{2}\left( \mathbf{n}\right) \delta _{ij}-%
\bar{b}_{i}\left( \mathbf{n}\right) \bar{b}_{j}\left( \mathbf{n}\right)
\right) \mathcal{I}_{0}\left( \mathbf{n,}\omega \right) \right\rangle
\label{PiAo}
\end{equation}%
In the case of $M_{J}=0$ when $\mathbf{\bar{b}=b}_{0}$, from Eqs. (\ref{Ta1}%
), (\ref{Ta2}) and (\ref{BA0}), (\ref{e}) we get 
\begin{equation}
\mathbf{\hat{T}}^{\left( 1\right) }=\sqrt{\frac{3}{2}}\frac{\omega \Delta }{%
\left( \omega +i0\right) ^{2}-\Delta ^{2}/5}\left[ \mathbf{e}^{\ast }\left( 
\mathbf{\hat{\sigma}b}_{1}\right) i\hat{\sigma}_{2}+\mathbf{e}\left( \mathbf{%
\hat{\sigma}b}_{-1}\right) i\hat{\sigma}_{2}\right] ~,  \label{T1f}
\end{equation}%
\begin{equation}
\mathbf{\hat{T}}^{\left( 2\right) }=\sqrt{\frac{3}{2}}\frac{\omega \Delta }{%
\left( \omega +i0\right) ^{2}-\Delta ^{2}/5}\left[ i\hat{\sigma}_{2}\left( 
\mathbf{\hat{\sigma}b}_{1}\right) \mathbf{e}^{\ast }+i\hat{\sigma}_{2}\left( 
\mathbf{\hat{\sigma}b}_{-1}\right) \mathbf{e}\right] .  \label{T2f}
\end{equation}%
Poles of the vertex function correspond to collective eigen modes of the
system (see, e.g. \cite{Leinson2012,Leinson2010a,Leinson2011}). Thus, the
pole at $\omega ^{2}=\Delta ^{2}/5$ signals the existence of collective
oscillations of the total angular momentum. The pole location on the complex 
$\omega$-plain is chosen so as to obtain a retarded vertex. 

Principally, the decay of these collective
oscillations into neutrino pairs is also possible by giving the additive
contribution into neutrino energy losses via the axial channel of weak
interactions. Later we will return to this problem. Here we concentrate on
the PBF processes. In this case we are interested in $\omega >2\Delta \bar{b}%
\left( \theta \right) \geq \sqrt{2}\Delta $, and a small term $\Delta
^{2}/5\ll $ $\omega ^{2}$ in the denominator of Eqs. (\ref{T1f}) and (\ref%
{T2f}) can be discarded to obtain simpler expressions%
\begin{equation}
\mathbf{\hat{T}}^{\left( 1\right) }\left( \mathbf{n}\right) =\sqrt{\frac{3}{2%
}}\frac{\Delta }{\omega }\left[ \mathbf{e}^{\ast }\left( \mathbf{\hat{\sigma}%
b}_{1}\right) i\hat{\sigma}_{2}+\mathbf{e}\left( \mathbf{\hat{\sigma}b}%
_{-1}\right) i\hat{\sigma}_{2}\right] ~,  \label{T1ff}
\end{equation}%
\begin{equation}
\mathbf{\hat{T}}^{\left( 2\right) }\left( \mathbf{n}\right) =\sqrt{\frac{3}{2%
}}\frac{\Delta }{\omega }\left[ i\hat{\sigma}_{2}\left( \mathbf{\hat{\sigma}b%
}_{1}\right) \mathbf{e}^{\ast }+i\hat{\sigma}_{2}\left( \mathbf{\hat{\sigma}b%
}_{-1}\right) \mathbf{e}\right] .  \label{T2ff}
\end{equation}

Substituting expressions (\ref{T1ff}) and (\ref{T2ff}) in the second line of
Eq. (\ref{PiA}) we obtain the anomalous part of the axial polarization
tensor in the long-wave limit%
\begin{equation}
\left( \Pi _{ij}^{A}\right) _{\mathsf{anom}}=3\frac{p_{F}m^{\ast }}{\pi ^{2}}%
\left\langle \left( \delta _{ij}-\delta _{i3}\delta _{j3}\right) \bar{b}%
^{2}\left( \mathbf{n}\right) \mathcal{I}_{0}\left( \mathbf{n,}\omega \right)
\right\rangle .  \label{PiAa}
\end{equation}

Summing together the contributions, given in Eqs. (\ref{PiAo}) and (\ref%
{PiAa}), we obtain the complete response function in the axial channel: 
\begin{align}
\Pi _{ij}^{A}\left( M_{J}=0\right) & =-4\frac{p_{F}m^{\ast }}{\pi ^{2}}%
\left\langle \left[ \bar{b}^{2}\delta _{ij}-\bar{b}_{i}\bar{b}_{j}\right] 
\mathcal{I}_{0}\right\rangle  \notag \\
& +3\frac{p_{F}m^{\ast }}{\pi ^{2}}\left\langle \left( \delta _{ij}-\delta
_{i3}\delta _{j3}\right) \bar{b}^{2}\mathcal{I}_{0}\right\rangle .
\label{PIA}
\end{align}%
The imaginary part of the function $\mathcal{I}_{0}\left( \mathbf{n,}\omega
\right) $ arises from the poles of the integrand in Eq. (\ref{FFq0}) at $%
\omega =\pm 2E$:%
\begin{equation}
\bar{b}^{2}\left( \mathbf{n}\right) \mathrm{Im}\mathcal{I}_{0}\left( \mathbf{%
n,}\omega \right) =\Theta \left( \omega -2\Delta _{\mathbf{n}}\right) \frac{%
\pi \Delta _{\mathbf{n}}^{2}}{2\omega \sqrt{\omega ^{2}-4\Delta _{\mathbf{n}%
}^{2}}}\tanh \frac{\omega }{4T}.  \label{IFFq0}
\end{equation}

Using Eq. (\ref{IFFq0}) and Eqs. (\ref{imP}), (\ref{PI00}), and (\ref{PIA})
we obtain the imaginary part of the weak polarization tensor for the $^{3}$P$%
_{2}\left( M_{J}=0\right) $ superfluid neutron liquid 
\begin{gather}
\mathrm{Im}\Pi _{\mu \nu }\left( \omega >0\right) =-\delta _{\mu i}\delta
_{\nu j}C_{A}^{2}p_{F}m^{\ast }\frac{2}{\pi }\frac{1}{\omega }\tanh \frac{%
\omega }{4T}  \notag \\
\times \int \frac{d\mathbf{n}}{4\pi }\left[ \delta _{ij}-\frac{\bar{b}_{i}%
\bar{b}_{j}}{\bar{b}^{2}}-\frac{3}{4}\left( \delta _{ij}-\delta _{i3}\delta
_{j3}\right) \right] \frac{\Delta _{\mathbf{n}}^{2}\Theta \left( \omega
-2\Delta _{\mathbf{n}}\right) }{\sqrt{\omega ^{2}-4\Delta _{\mathbf{n}}^{2}}}%
.  \label{ImPi}
\end{gather}

\section{PBF neutrino energy losses}

\label{sec:PBF}

Now we substitute the obtained weak polarization tensor to Eq. (\ref{QQQ})
for the neutrino emissivity. Contraction of the tensor (\ref{ImPi}) with $%
\left( K^{\mu }K^{\nu }-K^{2}g^{\mu \nu }\right) $ gives: 
\begin{gather}
\mathrm{Im}\Pi _{\mu \nu }\left( \omega \right) \left( K^{\mu }K^{\nu
}-K^{2}g^{\mu \nu }\right) =-C_{A}^{2}\frac{p_{F}m^{\ast }}{\pi }\frac{1}{%
\omega }\tanh \frac{\omega }{4T}  \notag \\
\times \int \frac{d\mathbf{n}}{4\pi }\left[ 2\left( \omega ^{2}-k_{\parallel
}^{2}\right) -k_{\perp }^{2}\right] \frac{\Delta _{\mathbf{n}}^{2}\Theta
\left( \omega -2\Delta _{\mathbf{n}}\right) }{\sqrt{\omega ^{2}-4\Delta _{%
\mathbf{n}}^{2}}},  \label{Ic}
\end{gather}%
where we denote 
\begin{equation}
k_{\parallel }^{2}=\frac{1}{\bar{b}^{2}}\left( \mathbf{k\bar{b}}\right)
^{2}~,~k_{\perp }^{2}=k^{2}-k_{\parallel }^{2}.  \label{qq}
\end{equation}%
After some algebra we find the neutrino emissivity in the form:

\begin{equation}
Q\simeq \frac{2}{15\pi ^{5}}G_{F}^{2}C_{A}^{2}\mathcal{N}_{\nu }p_{F}m^{\ast
}T^{7}\int \frac{d\mathbf{n}}{4\pi }\frac{\Delta _{\mathbf{n}}^{2}}{T^{2}}%
\int_{0}^{\infty }dx\frac{z^{4}}{\left( 1+\exp z\right) ^{2}}~,  \label{eps}
\end{equation}%
where $\Delta _{\mathbf{n}}^{2}\equiv \Delta ^{2}\,\bar{b}^{2}\left( \mathbf{%
n}\right) =\frac{1}{2}\Delta ^{2}\left( 1+3\cos ^{2}\theta \right) $, and $z=%
\sqrt{x^{2}+\Delta _{\mathbf{n}}^{2}/T^{2}}$.

It is necessary to notice that a definition of the gap amplitude is
ambiguous in the literature. For example, in the case of $M_{J}=0$, our gap
amplitude is $\sqrt{2}$ times larger than the gap amplitude in Ref. \cite%
{Yakovlev1999} (denote it $\Delta _{\text{\texttt{YKL}}}$), where it is
defined by the relation $\Delta _{\mathbf{n}}^{2}=\Delta _{\text{\textsc{YKL}%
}}^{2}\left( 1+3\cos ^{2}\theta \right) $. However, the total anisotropic
gap $\Delta _{\mathbf{n}}$ entering the energy of the quasiparticles is the
same in both calculations, since $\Delta /\sqrt{2}=\Delta _{\text{\texttt{YKL%
}}}$.

Returning to the standard physical units we get \cite{Leinson2010} 
\begin{eqnarray}
Q &=&\frac{4G_{F}^{2}p_{F}m^{\ast }}{15\pi ^{5}\hbar ^{10}c^{6}}\left(
k_{B}T\right) ^{7}\mathcal{N}_{\nu }R  \notag \\
&=&1.170\,\times 10^{21}\frac{m^{\ast }}{m}\frac{p_{F}}{mc}T_{9}^{7}\mathcal{%
N}_{\nu }R~\ \frac{\mathsf{erg}}{\mathsf{cm}^{3}\mathsf{s}}~.  \label{Q}
\end{eqnarray}%
Remind that $G_{F}$ is the Fermi coupling constant, $C_{A}\simeq 1.26$ is
the axial-vector weak coupling constant of a neutron, and $\mathcal{N}_{\nu
}=3$ is the number of neutrino flavors; $p_{F}$ is the Fermi momentum of
neutrons, $m^{\ast }\equiv p_{F}/v_{F}$ is the effective neutron mass; $m$
is bare nucleon mass, $T_{9}=T/(10^{9}\mathsf{K})$, $k_{B}$ is the Boltzmann
constant, and%
\begin{equation}
R=\frac{1}{2}C_{A}^{2}F_{t}.  \label{R}
\end{equation}%
The function $F_{t}$ is given by%
\begin{equation}
F_{t}=\int \frac{d\mathbf{n}}{4\pi }y^{2}\int_{0}^{\infty }dx\frac{z^{4}}{%
\left( 1+\exp z\right) ^{2}}.  \label{F}
\end{equation}%
Here the notation is used$\ z=\sqrt{x^{2}+y^{2}}$ with $y=\Delta _{\mathbf{n}%
}/T$.~The unit vector $\mathbf{n=p}/p$ defines the polar angles $\left(
\theta ,\varphi \right) $ on the Fermi surface.

It is necessary to stress that Eq.(\ref{eps}) as well as Eq. (\ref{Q})
involves the anomalous contributions into both the channels of weak
interactions (vector and axial). A comparison of the formula (\ref{R}) with
Eq. (28) of the work \cite{Yakovlev1999}, where the PBF neutrino losses were
obtained ignoring the anomalous interactions, allows one to see that the
anomalous contributions not only completely suppress the vector channel of
weak interactions, but also suppress four times the energy losses through
the axial channel. The resulting reduction of the emissivity of the PBF
processes in neutron matter is \cite{Leinson2010}:
\begin{equation}
\frac{C_{A}^{2}}{2\left( C_{V}^{2}+2C_{A}^{2}\right) }\simeq 0.19.  \label{q}
\end{equation}%
\begin{figure}[h]
\includegraphics{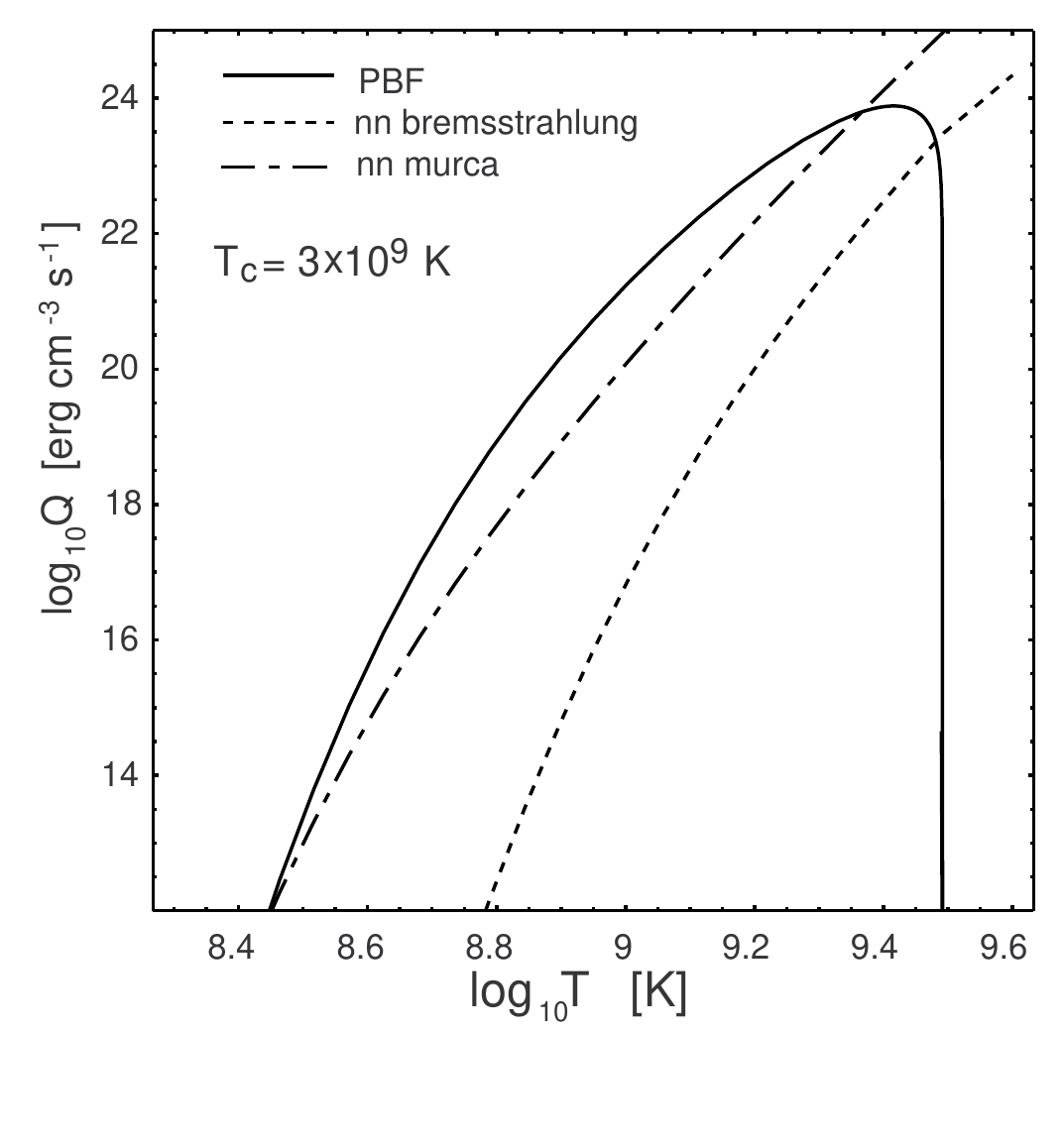}
\caption{PBF neutrino emissivity versus temperature $T$ in comparison with
the modified Urca and bremsstrahlung emissivities at $k_{F}=1.7$.}
\label{fig3}
\end{figure}
In spite of the so strong reduction, the neutrino emissivity caused by the
PBF processes can be the most powerful mechanism of the energy losses from
the NS core below the critical temperature $T_{c}$. In Fig. \ref{fig3}, the
PBF neutrino emissivity, as given in Eq. (\ref{eps}), is shown together with
the emissivities of modified Urca processes and bremsstrahlung multiplied by
the corresponding suppression factors resulting from superfluidity, as
obtained in Ref. \cite{YKGH}.

The emissivity from the PBF dominates everywhere below the critical
temperature for the $^{3}$P$_{2}$ superfluidity except the narrow
temperature domain near the critical point, where the modified Urca
processes are more operative.

\section{Decay of the eigenmodes of the condensate}

\label{sec:decay}

We now turn to an estimate of the neutrino energy losses due to decay of
thermally excited oscillations of the spin-triplet condensate of neutrons.
These eigenmodes represent collective oscillations of the direction of total
angular momenta of Cooper pairs which generate fluctuations of axial
currents in the superfluid system (spin density fluctuations). The energy of
the collective mode excitation $\omega =\Delta /\sqrt{5}$ is smaller than
the energy gap in the quasiparticle spectrum. In this case the function $%
\mathcal{I}_{0}\left( \mathbf{n,}\omega \right) $, given in Eq. (\ref{FFq0}%
), is real, and the imaginary part of the axial polarization tensor (\ref%
{PiA}) arises from the pole part of the functions $\mathbf{\hat{T}}^{\left(
1,2\right) }$ at $\omega ^{2}-\Delta ^{2}/5=0$.

With the aid of Sokhotsky's formula, $\left( \chi +i0\right) ^{-1}=\mathcal{P%
}\left( 1/\chi \right) -i\pi \delta \left( \chi \right) $, from the second
line of Eq. (\ref{PiA}) we get 
\begin{align}
\mathrm{Im}\Pi _{\mu \nu }\left( \omega >0\right) & =-\delta _{\mu i}\delta
_{\nu j}\left( \delta _{ij}-\delta _{i3}\delta _{j3}\right)  \notag \\
& \times \frac{3}{2\pi }C_{A}^{2}p_{F}m^{\ast }\left\langle \bar{b}^{2}%
\mathcal{I}_{0}\right\rangle \omega \delta \left( \omega -\Delta /\sqrt{5}%
\right) .  \label{PiS}
\end{align}%
The neutrino luminosity per unit volume is proportional to the product of
the total phase volume available to the outgoing neutrinos and the total
energy of the neutrino pair. This explains the temperature dependence of the
PBF neutrino emissivity, as given in Eq. (\ref{eps}). The presence of the
delta-function $\delta \left( \omega -\Delta /\sqrt{5}\right) $ in Eqs. (\ref%
{PiS}) restricts the total energy of the neutrino pair by the dispersion
relation and thus substantially reduces the total volume available to
neutrino pairs in the phase space. Integration over the phase volume will
result to appearance of the factor $\left( \Delta /\sqrt{5}\right) ^{7}$
instead of $T^{7}$. Just below the superfluid transition temperature, where
the main splash of the PBF neutrino emission occurs, the collective mode
energy $\omega _{s}=\Delta \left( T\right) /\sqrt{5}$ is small as compared
to the temperature. As a result the emissivity due to the collective mode
decays is many orders of magnitude slower than the PBF emissivity.

One might expect the two emissivities become comparable at sufficiently low
temperature $T\lesssim \Delta \left( T\right) /\sqrt{5}$. It is necessary to
notice, however, that our estimate is valid only when the anisotropic energy
gap is replaced by its average value in the anomalous vertices. Such an
approximation is good for the PBF processes but not for the eigen modes. The
exact account of the anisotropy dramatically reduces the neutrino losses due
to the collective mode decays \cite{Leinson2013}.

\section{Application to cooling modeling of neutron stars}

\label{sec:model}

The strong suppression of the vector PBF channel is basically incorporated
in the cooling simulations codes (e.g., \cite%
{Gupta2007,Page2009,Shternin2015,Potekhin2015,Han2017}). In the case of 
$^{1}$S$_{0}$ pairing of neutrons the suppression of the vector channel should be important in the cooling interpretation of a NS crust as the cooling time-scale of the crust is sensitive to the rates of neutrino emission.
Quenching of the neutrino emission, found in the case of $^{1}S_{0}$ pairing,
leads to higher temperatures that can be reached in the crust of an
accreting NS. This allows one to explain the observed data of
superbursts triggering \cite{Cumming2006,Gupta2007,Keek2008,Brown2009},
which was in dramatic discrepancy with the previous theory of the crust
cooling. However, the suppression of the neutron $^{1}$S$_{0}$ PBF process
does not lead to a distinguishable effect in the long-term cooling ($>$ 1000
years) of the star \cite{Page2009}.

The neutron pairing in the NS core, is expected to occurs into the spin-triplet $^{3}$P$_{2}$ state (a small $^{3}$F$_{2}$ admixture caused by tensor forces is normally neglected). Just a few years ago, suppression of the PBF neutrino emission due to spin-triplet neutron pairing in the NS core was included in the neutron star cooling codes only by complete suppression of the vector channel, while the emission in the axial vector channel remained unchanged \cite{Page2009,Page2011}. This corresponds to the reduction factor of $0.76$ with respect to the PBF emissivity previously obtained in  \cite{Yakovlev1999}, which led the authors to the conclusion that, within the minimal  cooling paradigm, the closing of the vector channel of the PBF neutrino emission does not significantly affect the long-term cooling of NSs. The reason is that the long-term cooling is controlled by
the axial channel of the PBF emissivities.

The suppression factor for PBF neutrino radiation given in Eq. (\ref{q})
involves two physical phenomena: (i) total suppression of the vector
channel, and (ii) the fourfold suppression of the axial channel caused by
the anomalous weak interactions. For the first time the suppression of the axial PBF channel was implemented in a simulation of the Cas A NS cooling in \cite{Shternin2011,Elshamouty2013}. 
It was found that the whole set of observations is quite consistent with the theoretical suppression factor of $ 0.19 $. This factor, presented in Eq. (\ref{q}), is now commonly used for suppression of the PBF reactions in spin-triplet  superfluid neutron matter of the NS cores 
(e.g. \cite{Shternin2015,Potekhin2015,Beloin2018,Fortin2018}).

An exhaustive numerical analysis of the anomalous axial PBF contribution to
the temporal evolution of the NS cooling is presented in \cite{Potekhin2018}.
The interested reader can get a clear idea about importance of this
contribution from Figs. 2 and 3 of that work, where the authors present the
NS cooling curves for the cases with and without the anomalous contribution.

\section{Conclusion}

\label{sec:concl}

We have discussed the important role of anomalous weak interactions in
mechanisms of neutrino emission taking place in fermionic superfluids
typical for the NS cores. It is established that due to the anomalous
contributions the PBF neutrino emissivity from the vector channel is almost
completely suppressed and can be ignored. This result is in agreement with
the conservation of vector current in weak interactions. In the case of
spin-singlet pairing the neutrino emission through the axial-vector channel
is also suppressed because the total spin of the Cooper pair $S=0$ is
conserved in the non-relativistic case. Thus the neutrino energy losses due
to singlet-state pairing of baryons can, in practice, be ignored in
simulations of NS cooling. This makes unimportant the PBF neutrino losses
from pairing of protons or hyperons.

The minimal cooling paradigm assummes that the direct Urca processes and any exotic fast reactions are not operative in the NC core. In this scenario, neutrino emission at the long-term cooling epoch comes mainly from modified Urca processes, nn-bremsstrahlung, and from the "PBF" processes, which arise in the presence of spin-triplet superfluidity of neutrons \cite{Page2009}. We have shown that the anomalous weak interactions in the $^{3}$P$_{2}$ superfluid suppress the PBF neutrino emission, although not so sharply as in spin-singlet superfluid liquids. Namely, the vector channel of weak interactions is again strongly suppressed and can be ignored while the neutrino losses through the axial channel are suppressed only partially. Despite of the approximately fivefold total suppression, the PBF mechanism of the neutrino energy losses is still operative.
In many cases, especially for temperatures near the critical superfluidity temperature of neutrons, the PBF neutrino reactions can dominate and should be accurately taken into account.

\end{document}